# Universal low-temperature Ohmic contacts for quantum transport in transition metal dichalcogenides

Shuigang Xu[1†], Zefei Wu[1†], Huanhuan Lu[1†], Yu Han[1], Gen Long[1], Xiaolong Chen[1], Tianyi Han[1], Weiguang Ye[1], , Yingying Wu[1], Jiangxiazi Lin[1], Junying Shen[1], Yuan Cai[1], Yuheng He[1], Fan Zhang[2], Rolf Lortz[1], Chun Cheng[3], Ning Wang[1*]

[1]Department of Physics and Center for 1D/2D Quantum Materials, the Hong Kong University of Science and Technology, Clear Water Bay, Hong Kong, China

[2]Departement of Physics, the University of Texas at Dallas, Richardson, Texas 75080, USA

[3]Department of Materials Science and Engineering and Shenzhen Key Laboratory of Nanoimprint Technology, South University of Science and Technology, Shenzhen 518055, China

[†]These authors contributed equally

[*]Corresponding author: phwang@ust.hk

# Abstract

Low carrier mobility and high electrical contact resistance are two major obstacles prohibiting explorations of quantum transport in TMDCs. Here, we demonstrate an effective method to establish low-temperature Ohmic contacts in boron nitride encapsulated TMDC devices based on selective etching and conventional electron-beam evaporation of metal electrodes. This method works for most extensively studied TMDCs in recent years, including $MoS_2$, $MoSe_2$, $WSe_2$, $WS_2$, and 2H-$MoTe_2$. Low electrical contact resistance is achieved at 2 K. All of the few-layer TMDC devices studied show excellent performance with remarkably improved field-effect mobilities ranging from 2300 cm$^2$/V s to 16000 cm$^2$/V s, as verified by the high carrier mobilities extracted from Hall effect measurements. Moreover, both high-mobility n-type and p-type TMDC channels can be realized by simply using appropriate contact metals. Prominent Shubnikov–de Haas oscillations have been observed and investigated in these high-quality TMDC devices.

**Introduction**

Experimental studies on the electrical transport properties of atomically thin semiconducting transition metal dichalcogenides (TMDCs) [1-9] have encountered a number of obstacles, such as high impurity, low carrier mobility, and high electrical contact resistance caused by Schottky barriers formed at the metal-TMDC interfaces [10, 11]. Ohmic contacts between metals and TMDCs are difficult to achieve because of work function mismatches and Fermi level pinning effects [12]. Over the last few years, significant efforts have attempted to improve the quality of electrical contacts in metal-TMDCs, such as through using work function-matched metals [13], vacuum annealing [14], and phase engineering [15]. The quality of electrical contacts at metal-$WSe_2$ interfaces, for example, have been gradually improved by surface doping treatment [11] and utilization of different materials with low (or tunable) work functions [10, 16]. Electrical contact problems (mainly due to Schottky barriers) exist in all semiconducting TMDC devices contacted using metal electrodes particularly at cryogenic temperatures. They significantly limit the injection of charge carriers into the transport channel and thus prohibit detection of quantum transport properties in these 2D materials. Reducing the Schottky barrier at metal-TMDC interfaces is the most critical step in advancing the device transport performance of TMDCs. Recently, ultrahigh electron mobility has been achieved in $MoS_2$ devices constructed based on graphene multi-terminal electrodes [17]. The contact resistance in these graphene-terminated devices can be largely reduced by applying a large back gate voltages, and quantum oscillations have been detected which show complicated multi-band features.

Based on our systematic investigations of electrical contacts in various TMDCs, we find that cleanliness of TMDC interfaces is the critical issues in realizing high-quality Ohmic contacts and fabricating high performance devices. To eliminate the contamination/oxidation occurring on TMDC surfaces, we use a dry transfer technique in an inert environment of argon or nitrogen and protect few-layer TMDCs using ultrathin hexagonal boron nitride (h-BN) sheets [18]. Ultrathin h-BN is well-known dielectric material that provides a charge-free environment for high-quality graphene devices [19, 20]. Ultrathin h-BN is also very effective in blocking charge impurities from the $SiO_2$ substrate and hence increasing carrier mobility in TMDCs, particularly at low temperatures. To achieve high-quality Ohmic contacts, we develop a selective etching process to open windows from the top h-BN cover layer and deposit metal electrodes using standard electron-beam (e-beam) lithography and conventional e-beam evaporation techniques. This method is different from the edge contact configuration [20] of h-BN encapsulated graphene. The electrical Ohmic contacts prepared on the top surfaces of TMDCs by this simple etching process are very stable and can survive in air for several months (see figure S12 in Supporting Information). High field-effect mobilities of up to 16000 $cm^2$/V s (Hall mobilities of up to 9900 $cm^2$/V s) have been routinely achieved in n-type TMDCs. Furthermore, high mobility p-type TMDCs channels are achieved using Pd contacts since Pd has a high work function and can match the valence bands of TMDCs. As an example, high quality p-type $WSe_2$ with hole mobility of up to 8550 $cm^2$/V s is demonstrated. The fabrication of high quality metal electrodes for atomically thin TMDCs demonstrated here is more practical and reliable since the contact resistance weekly depends on applied gate voltages. Remarkable improvements in contact resistance and carrier mobility allow us to observe prominent quantum phenomena, such as Shubnikov–de Haas (SdH) oscillations, in good agreement with the intrinsic properties of the band structures in few-layer TMDCs.

**Results and discussion**

**1. Fabrication of metal contacts for h-BN encapsulated TMDCs**

Figure 1 schematically illustrates the fabrication of a field-effect transistor (FET) device from a few-layer TMDC encapsulated by h-BN sheets. The high-quality single crystalline TMDCs used in this work are either obtained commercially (e.g., $MoS_2$) or grown via the chemical vapor transport method (e.g., $WSe_2$, $WS_2$, $MoSe_2$, and $2H-MoTe_2$) [21]. Few-layer TMDC samples are mechanically exfoliated on a Si substrate coated with 300 nm-thick $SiO_2$. To eliminate possible contamination, a polymer-free dry transfer technique recently developed for device fabrication of graphene is adopted in this study [20]. A few-layer TMDC is picked up from the $SiO_2$/Si substrate by an ultrathin h-BN flake through van der Waals interactions. The h-BN/TMDC flake is then transferred to a fresh h-BN flake previously exfoliated on another $SiO_2$/Si substrate to form a h-BN/TMDC/h-BN sandwich structure. A high temperature annealing process (under $H_2$/Ar atmosphere) is necessary to remove the small bubbles formed at the interfaces between h-BN and TMDCs. Although edge contacts can be established for our samples, the contact resistance is normally high, resulting in difficulties in detecting the intrinsic properties of TMDCs (see Supporting Information). The poor performance of the devices with edge contacts indicates that even TMDCs are encapsulated by h-BN sheets, if the contact problems are not solved, the intrinsic properties of the TMDC devices cannot be detected. By contrast, the etching process we developed (etching the top of h-BN only) can largely reduce the TMDC contact resistance. The contact areas are patterned by standard e-beam lithography techniques, and the exposed top h-BN layer is subsequently etched by $O_2$ plasma at a relatively high speed. By controlling the etching duration, TMDCs can be partially exposed for metal deposition because the etching rate of TMDCs by $O_2$ plasma is low.

## 2. Low-temperature Ohmic contacts for TMDC channel materials

To demonstrate the Ohmic contact characteristics of our TMDC devices, $I_{ds}$-$V_{ds}$ measurements are first carried out in a two-terminal configuration at both room and cryogenic temperatures. Figures 2(a)-(c) display typical examples of the $I_{ds}$–$V_{ds}$ curves measured under different back gates ($V_g$) for few-layer n-type $WSe_2$, p-type $WSe_2$, and n-type $MoSe_2$, respectively. Excellent Ohmic contact characteristics are maintained at 2 K. In our TMDC samples, the linear characteristics of the $I_{ds}$–$V_{ds}$ curves are held from 2 K to 300 K (see figure S3 in Supporting Information). Moreover, the linearity and zero Schottky barrier height extracted from the Arrhenius plot (see figure S5 in Supporting Information) can be held at different back gate voltages (figures 2(a)-(c)). Different from graphene-TMDCs contacts [22-24], the metal contact resistance fabricated by the present etching process is weakly dependent of gate voltages.

By comparing the electrical contact characteristics of two- and four-terminal configurations, the contact resistance is extracted according to [25, 26] $R_c = R_{2p} - \frac{L_{tot}}{L_{in}} R_{4p}$, where $L_{tot}$ and $L_{in}$ are the total and inner channel lengths, respectively. For n-type $WSe_2$, the contact resistance is only 50 Ω at $V_g$=35 V and saturates to 150 Ω at high gate voltages (see figure S9 in Supporting Information). Given that the width of the channel is 7.0 µm and that both source and drain contributions are considered, the unit length contact resistance is within the range of 0.2–0.5 kΩ·µm, which is considerably lower than those achieved in ionic liquid gating graphene/$WSe_2$ (2 kΩ·µm) [10] and in Ag/$WSe_2$ (6.5 kΩ·µm) [16] contacts. The most efficient contact resistances achieved by MoSe$_2$ and MoS$_2$ are about 0.3 k$\Omega \cdot \mu$m (V$_g$=70 V) and 0.5 k$\Omega \cdot \mu$m (V$_g$=80 V), respectively. The low contact resistance in our samples is as efficient as those achieved in metal/MoS$_2$ contacts by nickel-etched graphene electrodes (0.2 k$\Omega \cdot \mu$m) [27] and by phase engineering (0.2–0.3 k$\Omega \cdot \mu$m) [15]. The low contact resistances reported by these work [15, 27] were achieved at room temperature. In the present study, the low contact resistances we achieved have been extended to cryogenic temperatures.

We believe that the excellent performance of the metal contacts we developed should not be due to doping effects although the surfaces of few-layer TMDC channels are etched more or less by the oxygen plasma. We have used Raman spectroscopy to characterize possible impurities which might be induced by the etching process (see figure S20 in Supporting Information). However, no doping related Raman shift can be detected [28]. This is consistent with the fact that the high mobility p-type WSe$_2$ devices with high-quality low temperature Ohmic contacts can be easily fabricated using Pd as the contact electrodes. If our oxygen plasma etched TMDC windows areas (for making the electrodes) are heavily n-type doped, it is not possible to realize the hole injection through the Pd electrodes since the Fermi level should be pinned at the conduction band. The control experiments shown in figures S6 and S7 in Supporting Information demonstrate that the quality of the contacts and thus the performance of the TMDC devices fabricated by the etching process are highly dependent on both bottom and top h-BN sheets. Obviously, the h-BN sheets play an important role in achieving Ohmic contacts in our TMDC samples.

We find that the contact barrier for electron injection is very low if Ti (low work function) is used as the contact metal. In this case, the barrier for the hole injection is very high and no hole injection is observed even under a large negative gate voltage. In contrast, if Pd (high work function) is used as the electrodes, the contact barrier is very low for hole (p-type) injection but very high for electron injection. In this case, we cannot observe any electron injection under a large positive gate voltage. For the devices fabricated on SiO$_2$/Si substrates, because the charge traps and disorder formed at the TMDCs/SiO$_2$ interface induce a large number of interfacial gap states and result in a high Schottky barrier with the Fermi level pinned to the band gap of the TMDCs at the contact region. Therefore, the barriers for both electron and hole injections are large but smaller than the bandgap. Therefore, the devices showed ambipolar behaviors when either Ti or Pd is used as the contact metals (see figures S13 and S14 in Supporting Information).

3. **Mobility characterization for TMDC devices**

Figure 3 shows the temperature-dependent transport characteristics of several TMDCs. The Ti contacted WSe$_2$ device shows n-type behaviors as revealed by the increasing conductance with increasing positive gate voltages, while the Pd contacted WSe$_2$ shows p-type behaviors. Under high gate voltages, the conductance (measured by the four-terminal configuration) dramatically increases at low temperatures. As demonstrated in a n-type WSe$_2$ sample, at V$_g$=70 V, the resistance dramatically decreases from 13720 Ω (at room temperature) to 290 Ω at 1.7 K.

Our few-layer TMDC devices exhibit excellent performance at low temperatures because of their barrier-free contacts. The field-effect mobility is extracted from the linear region of the conductance curves [4] using $\mu = [dG/dV_g] \times [\frac{L}{WC_g}]$, where L and W are the channel length and width, respectively, and $C_g$ is the gate capacitance (the thickness of SiO$_2$ is 300 nm). In this study, $C_g$ is extracted from the Hall effect measurements, which is considered a more accurate method [29, 30]. At 2K, the field-effect mobilities achieved in n-type WS$_2$, MoS$_2$, WSe$_2$, MoSe$_2$ and MoTe$_2$ are 16000, 14000, 8600, 4400 and 2300 cm$^2$/V s, respectively, which are summarized in table 1.

The etching process reported here is also very effective for making high quality Ohmic contact to semiconducting 2H phase MoTe$_2$ channels which have been reported to be unstable in air [31]. The 2H-MoTe$_2$ FET devices fabricated using our method show excellent performance with high mobility up to 2300 cm$^2$/V s at 2 K. The reproducibility of this kind of high-quality samples is acceptable (more than 15 high quality devices have been obtained). Table S1 in Supporting Information summarizes the performances of our TMDC devices. The present selective etching method is applicable to ultrathin TMDCs with thickness down to two layers indicating that the etching process can be controlled fairly well. Although the mobilities of these extremely thin samples show quality degradation, the mobilities we achieved in bilayer and trilayer (1200 cm$^2$/V s and 3900 cm$^2$/V s respectively), for example, are considerably high. We believe that minor defect states are introduced by the O$_2$ plasma and degrade the quality of ultrathin thin TMDC contact channels. However, compared with the impurities induced gap states from SiO$_2$, the damage or defects induced by O$_2$ plasma are insignificant. According to our investigation, the h-BN encapsulation structure can effectively reduce the charge traps in TMDC devices. The h-BN/TMDC/h-BN sandwich structures do not exhibit any hysteresis effect (see figure S2 in Supporting Information). However, pronounced hysteresis effects are observed in the WSe$_2$ devices directly prepared on SiO$_2$/Si substrates (see details in Supporting Information).

The temperature-dependent field-effect mobility and Hall mobility data of few-layer TMDC devices are plotted in log scale in figures 3(d)–3(g). At high temperature regions (e.g. above ~40 K for n-type WSe$_2$), the mobility is suppressed by phonon scattering, which can be fitted by $\mu \sim T^{-\gamma}$. The saturation of μ suggests impurity-dominated scattering at low temperatures. The Hall mobility is extracted according to $\mu_\mathrm{H} = \sigma/ne$, where $\sigma = GL/W$ is the sheet conductance, $n = 1/R_\mathrm{H}e$ is the carrier density, $R_H$ is the Hall coefficient, and $e$ is the charge of an electron. For n-type WSe$_2$, the room temperature $\mu_\mathrm{F}$ is 105 cm$^2$/V s and reaches 8600 cm$^2$/V s at T=2 K. The $\mu_\mathrm{H}$ can reach 7100 cm$^2$/V s for a carrier density of $5.1 \times 10^{12}$ cm$^{-2}$ at T=2 K. The temperature-dependent $\mu_\mathrm{H}$ shows a characteristic similar to that of $\mu_\mathrm{F}$ (figure 3(b)), with $\gamma = 1.9$ extracted from both $\mu_\mathrm{F}$ and $\mu_\mathrm{H}$. Similarly high mobilities are also observed in MoS$_2$ and MoSe$_2$. The phonon-related parameters γ in all of the TMDCs are generally larger than those reported previously [14, 30, 32]. We observe that $\mu_\mathrm{H}$ varies with the carrier densities. This is because when the carrier density is high, free electrons can screen the charge impurities [33] resulting in increasing $\mu_\mathrm{H}$ with increasing carrier density ($d\mu_\mathrm{H}/dn > 0$), as shown in the Supporting Information. The dependence of Hall mobility on carrier density also results in a larger field-effect mobility compared with the Hall mobility in the same sample (i.e., the Hall factor $r_\mathrm{H} > 1$) [14] because $\mu_\mathrm{Field} = \mu_\mathrm{H} + nd\mu_\mathrm{H}/dn$.

## 4. Shubnikov-de Haas oscillations in TMDCs

Because of the reliable low-temperature Ohmic contacts and high mobilities, the SdH oscillations of the magnetoresistance (MR) can be easily detected in our TMDC devices. Figures 4(a) and 4(e) show the $R_{xx}$ plotted as a function of magnetic fields at different gate voltages for few-layer p-type and n-type WSe$_2$, respectively. Pronounced oscillations are clearly visible in the MR data. These quantum oscillations arise from Landau quantization of the cyclotron motion of charge carriers [34]. They depend only on the perpendicular component of the applied magnetic field as verified by the field-angle dependence experiment shown in Supporting Information, revealing the 2D nature of charge carriers in few-layer WSe$_2$.

The temperature dependence of SdH oscillations in p-type few-layer WSe$_2$ at low magnetic fields is plotted in figure 4(b), in which a polynomial background is subtracted in order to extract the oscillation amplitude. At low magnetic fields in which the Zeeman effect is negligible, the SdH amplitude can be well described by the Lifshitz–Kosevich formula [35] $\Delta R(B,T) = R_D \lambda(B,T)/\sinh\lambda(B,T)$. $R_D = \exp(-2\pi^2 m^* k_B T_D/\hbar eB)$ is the Dingle factor, and $\lambda(B,T) = 2\pi^2 m^* k_B T/\hbar eB$ is the parameterized thermal damping, where $k_B$ is the Boltzman constant, $\hbar$ is the reduced Planck constant, and $m^*$ is the cyclotron effective mass. Figure 4(c) shows the fitting results of the SdH oscillations in a p-type WSe$_2$ sample, which yield $m^* = 0.44 m_0$, where $m_0$ is the rest mass of an electron. Similar analysis can yield $m^* = 0.16 m_0$ in a n-type WSe$_2$.

After extracting the effective mass, the quantum scattering time $\tau_q$ and quantum mobility $\mu_q$ can be accurately determined from the Dingle factor, where $\lambda_D = 2\pi^2 k_B m^* T_D / \hbar eB$. $T_D$ is the Dingle temperature, given by $T_D = \hbar / 2\pi k_B \tau_q$. By plotting $\ln[\Delta R(B,T)B\sinh\lambda(T)]$ as a function of $1/B$, the slope of the linear fitting curves (shown in figure 3(d)) will yield $T_D$. At T=2 K, the fitting results in the Dingle temperature $T_D = 4.5\,\mathrm{K}$. The corresponding hole quantum scattering time $\tau_q = 269\,\mathrm{fs}$ and the quantum mobility $\mu_q = e\tau_q/m^* = 1100\,\mathrm{cm}^2/\mathrm{Vs}$ in p-type WSe$_2$. For n-type WSe$_2$, $\tau_q = 404$ fs and $\mu_q = 4400$ cm$^2$/V s at $V_g$=50 V. For both of p-type and n-type samples, $\mu_q$ is slightly smaller than the corresponding Hall mobility, since the quantum scattering time is shorter than the transport scattering time ($\tau_q < \tau_t$). This result indicates that the dominant scattering in few-layer WSe$_2$ is due to the long-range scattering caused by charge impurities [17]. By further reducing interfacial impurities or under sufficiently high magnetic fields, the sample quality should be ready for the exploration of quantum Hall and other quantum phenomena in TMDCs [34].

By comparing the carrier density measured by Hall effects with carrier density obtained from oscillation periods, we find that the n-type and p-type few-layer WSe$_2$ show different behaviors in valley degeneracy. Figure 4(e) shows the SdH oscillations of an n-type few-layer WSe$_2$ sample under magnetic fields of up to 14 T. The oscillations can be observed more clearly in figure 4(f) by plotting the numerical derivative dR/dB as a function of 1/B. Fourier transformation of the oscillatory component reveals that, besides the first harmonic at B$_F$=20 T, a second harmonic appears near B$_F$=40 T. This second peak indicates the small spin-valley splitting between the Q-valleys along ΓK. The carrier density participating in the SdH oscillations, according to $n_s = g_s g_v e B_F / h$ ($g_s = 2$ for the spin degeneracy, and $g_v = 6$ for the Q-valley degeneracy), is approximately $5.8 \times 10^{12}$ cm$^{-2}$, which is in good agreement with the density ($6.0 \times 10^{12}$ cm$^{-2}$) measured by Hall effects. This further confirms that in few-layer WSe$_2$ [36] the conduction band bottom is located at the six-fold Q-valleys. For p-type few-layer WSe$_2$, the valance band top is located at the Γ-valley with $g_s = 2$ and $g_v = 1$. By considering the degeneracy in the Γ-valley, the calculated carrier density participating oscillations is $n_s = 2.8 \times 10^{12}$ cm$^{-2}$ at V$_g$=-60 V for p-type WSe$_2$, which is consistent with the corresponding carrier density measured by Hall effects ($2.7 \times 10^{12}$ cm$^{-2}$).

**Conclusion**

In summary, we have demonstrated a simple and effective approach for fabricating high-mobility atomically thin TMDC field-effect devices. High-quality low-temperature Ohmic contacts have been achieved via a selective etching process. We have found that clean interfaces and very low contact resistance realized in the h-BN/TMDC/h-BN encapsulation structures are critical in detecting the intrinsic transport properties of ultrathin TMDCs. Few-layer TMDC devices exhibit ultrahigh field-effect mobility at low temperatures, as well as clear quantum oscillations in intermedium magnetic field. The experimental results presented in this work pave the way for improving the quality of atomically thin TMDC devices and for exploring their quantum transport phenomena.

**Methods**

The h-BN encapsulated TMDC devices are fabricated by a dry transfer technique in an inert environment of argon or nitrogen. In a typical process, atomically thin TMDC sheets are mechanically exfoliated on $SiO_2$/Si substrates. The layer number of TMDC is determined by optical contrast, photoluminescence, and atomic force microscopy. Ultrathin h-BN sheets are exfoliated on a glass-supported PMMA film. We use a large h-BN sheet to pick up TMDC flakes through van der Waals interactions. The h-BN/TMDC stacks are then aligned to another h-BN thin sheet that had been previously exfoliated on $SiO_2$/Si substrates under an optical microscope. The PMMA-supported h-BN/TMDC is then transferred to the bottom h-BN to form a h-BN/TMDC/h-BN sandwich structure. PMMA is later removed by acetone. The final device structures are annealed under $Ar/H_2$ at 400 °C to remove small bubbles. To fabricate the metal contacts, the h-BN encapsulated TMDC is first patterned by standard e-beam lithography, and the top h-BN is etched by $O_2$ plasma via reaction-ion etching. The flow rate of $O_2$ is 40 sccm. The RF power is 200 W. Under this condition, the etching rate is about 10 nm/min for h-BN and 1 nm/min for TMDCs. For thick h-BN (>20 nm), the top h-BN can be initially rapidly etched by $O_2$ (4 sccm) mixed with $CHF_3$ (40 sccm) under a RF power of 200 W (the etching rate for h-BN in this recipe is about 100 nm/min) and then precisely etched by $O_2$ plasma only. Since the $O_2$ plasma etching rate for TMDC is low, the control of the etching process is not difficult. To guarantee the conducting channels are fully encapsulated by h-BN, a small overlap between the metal electrodes and the top h-BN is made by the second e-beam lithography for electrode patterning. Then, standard e-beam evaporation of Ti/Au (5 nm/80 nm) for n-type devices or Pd/Au (20 nm/80 nm) for p-type devices is carried out to deposit the electrodes on the etched patterns. To remove potential residues from PMMA, the samples are cleaned by $O_2$ plasma (8 s at 100 W) before metal electrode deposition. Characterization of electrical transport properties is conducted at cryostat temperatures, and the two-terminal I–V curves are measured by a Keithley 6430 instrument. Other transport data are collected by standard lock-in amplifier techniques.

**Acknowledgement**

The authors acknowledge the financial support provided by the Research Grants Council of Hong Kong (Project Nos. 16302215, HKU9/CRF/13G, 604112, and N_HKUST613/12) and the technical support provided by the Raith-HKUST Nanotechnology Laboratory for the electron-beam lithography facility at MCPF. F.Z. is supported by UT Dallas research enhancement funds, and thanks Gui-Bin Liu for helpful discussions.

**Additional information**

Supporting information is available.

**Competing financial interests**

The authors declare no competing financial interests.

**Author contributions**

S.G.X. and H.H.L. grew the WSe$_2$, MoSe$_2$ and MoTe$_2$ bulk crystal. S.G.X. and H.H.L. fabricated WSe$_2$ devices. Z.F.W. fabricated MoS$_2$ and WS$_2$ devices. S.G.X. fabricated MoSe$_2$ and MoTe$_2$ devices. S.G.X., Z.F.W., H.H.L. and Y.H. performed and analyzed the measurements under supervision of N.W.. S.G.X. and Y.H. performed the high magnetic field measurements with the help of J.Y.S. and R.L.. S.G.X., Z.F.W., F.Z., and N.W. analyze the quantum transport data. Other authors provided technical assistance on sample preparations and measurements and discussed on the paper. S.G.X. and N.W. wrote the paper with input from all authors.

2013 Optical signature of symmetry variations and spin-valley coupling in atomically thin tungsten dichalcogenides *Sci. Rep.* **3** 1608

Figures and Legends

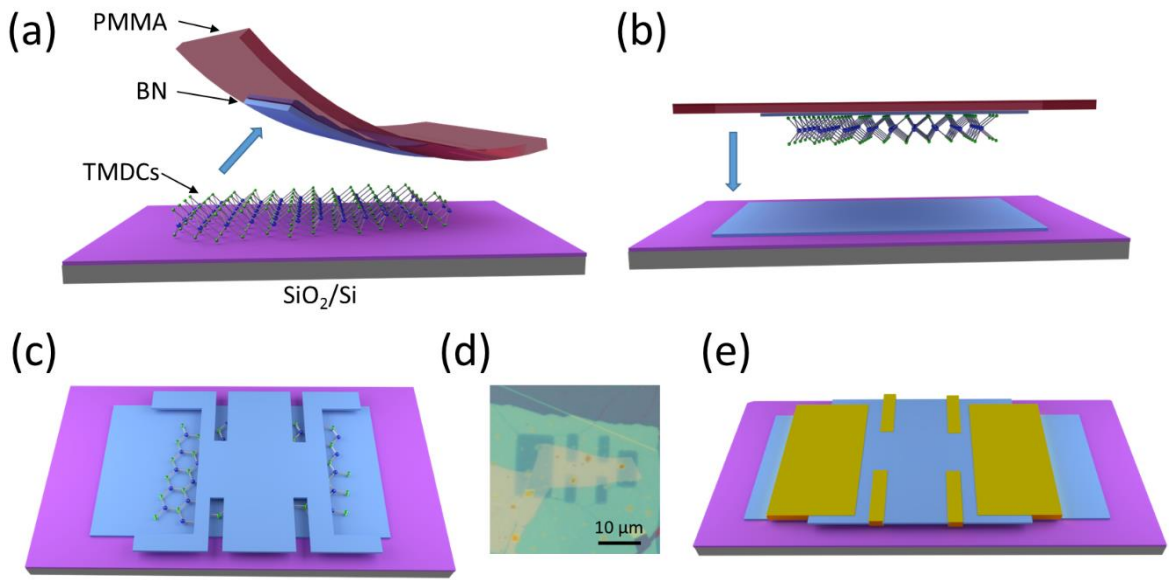

**Figure 1. Fabrication flow of h-BN encapsulated few-layer TMDC devices.** (a), Schematic of TMDC picked up by a h-BN flake on PMMA. (b), TMDC/h-BN is transferred to the bottom h-BN to form a h-BN/TMDC/h-BN sandwich structure. (c), Contact regions with a Hall bar pattern are etched by $O_2$ plasma. (d), Optical image of the device after $O_2$ plasma etching. (e), Final device obtained after deposition of metal electrodes.

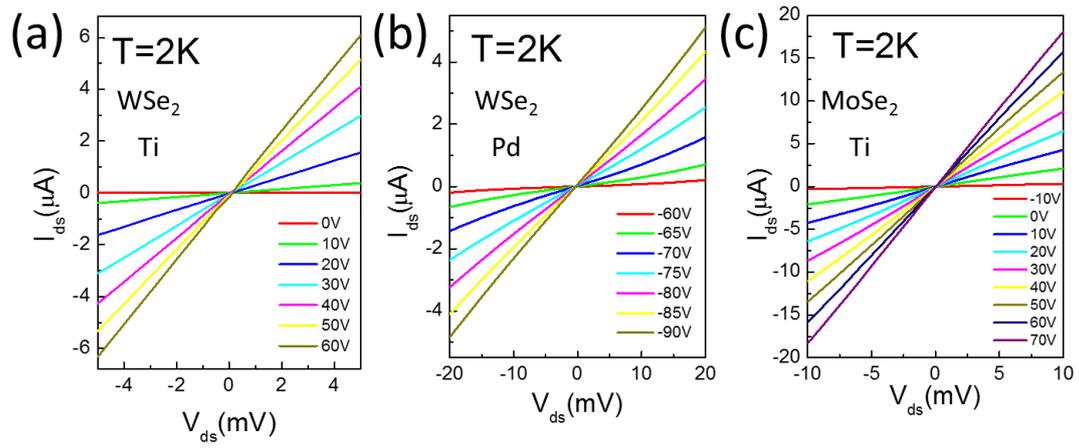

**Figure 2. Performance of the metal electrode contact in the h-BN/TMDC/h-BN van der Waals structures**. (a) –(c), $I_{ds}$-$V_{ds}$ characteristics obtained by a two-terminal configuration at T=2 K for (a) n-type $WSe_2$ (b) p-type $WSe_2$ and (c) n-type $MoSe_2$. Linear $I_{ds}$-$V_{ds}$ curves are observed at various back gates, indicating Ohmic contacts.

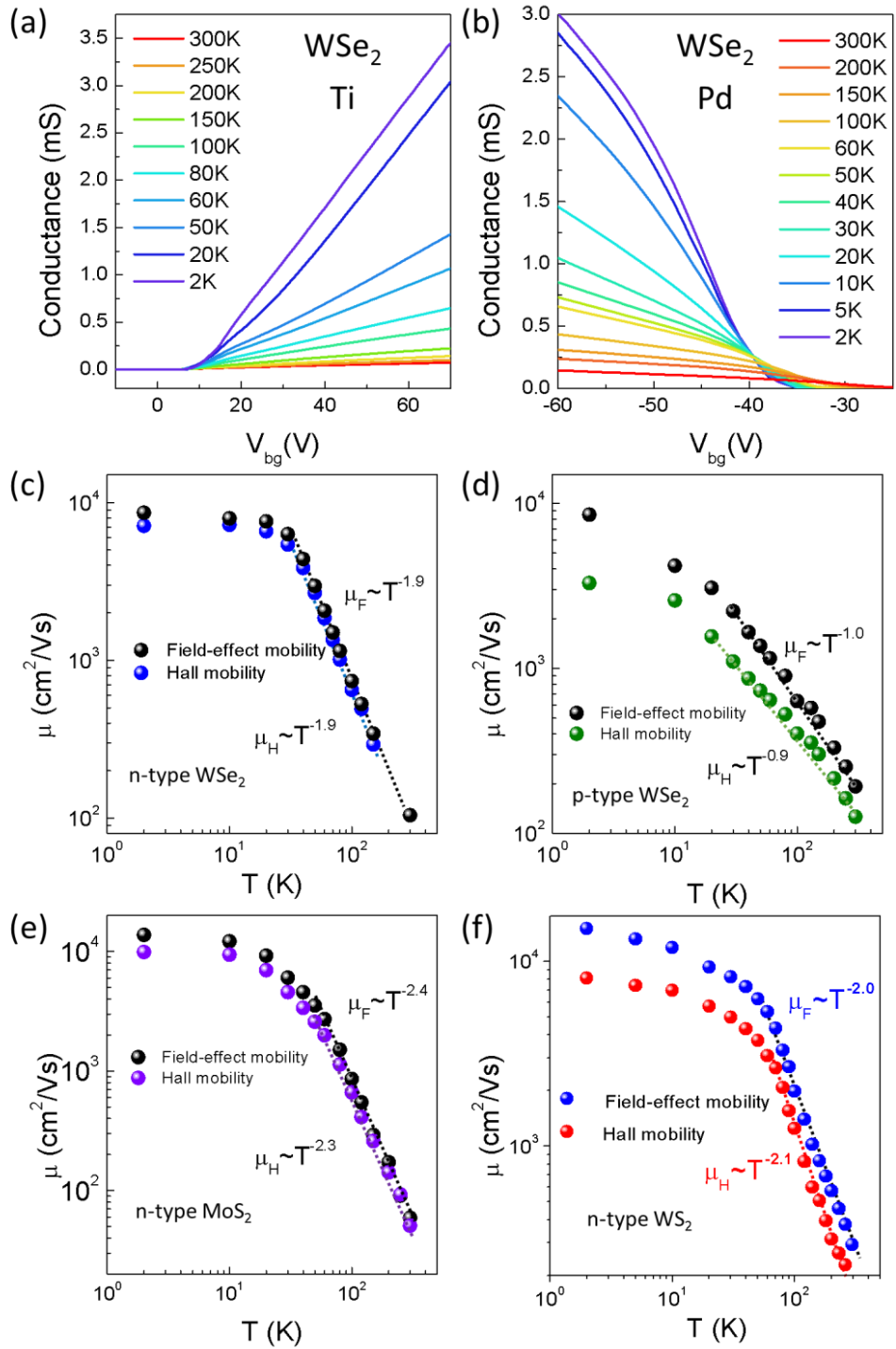

**Figure 3. Temperature-dependent conductance and mobility engineering.** (a), (b), Four-terminal conductance plotted as a function of back gate voltage at various temperatures: (a) n-type WSe$_2$ with Ti as contact metal; (b) p-type WSe$_2$ with Pd as contact metal. (c)-(f), Field-effect and Hall mobilities at various temperatures for few-layer (c) n-type WSe$_2$, (d) p-type WSe$_2$, (e) n-type MoS$_2$, and (f) n-type WS$_2$. Linear fittings were used in the phonon-limited region to extract γ.

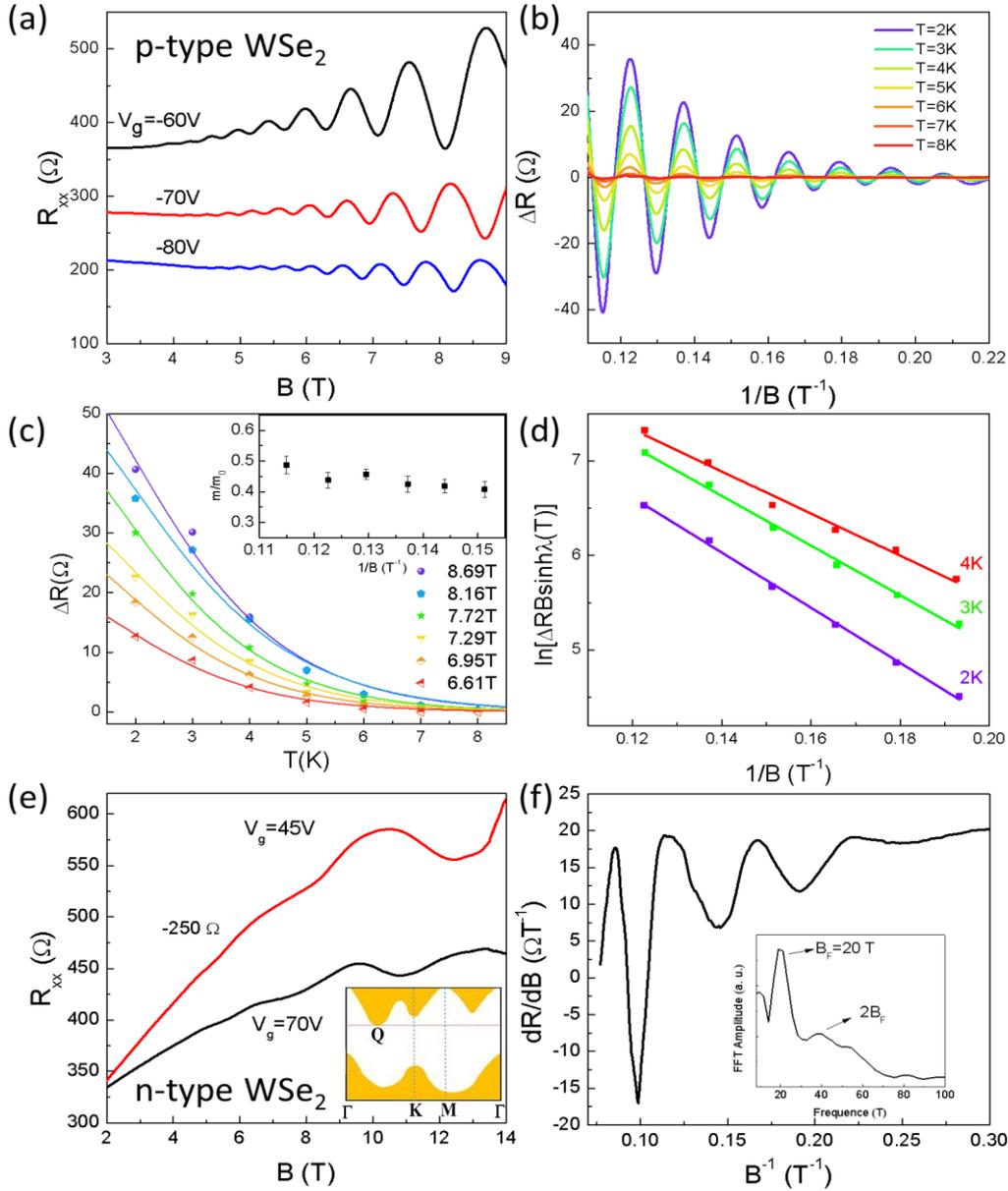

**Figure 4. Shubnikov–de Haas oscillations in h-BN encapsulated few-layer TMDCs**. (a), Longitudinal resistance plotted as a function of magnetic field at various voltages in a few-layer p-type $WSe_2$ sample. (b), SdH oscillation as a function of 1/B measured at different temperatures with $V_g$ fixed at -70 V. To obtain $\Delta R$, the magnetroresistance $R_{xx}$ is subtracted by a smooth background. (c), SdH oscillation amplitudes as a function of temperature at different magnetic fields. The solid lines are the fitting results using the Lifshitz-Kosevich formula. The inset shows the fitting results of the cyclotron effective mass. (d), Dingle plots of $\ln[\Delta R \cdot B \cdot T \cdot B \cdot \sinh\lambda]$ as a function of 1/B at different temperatures. Quantum scattering time and quantum mobility can be extracted from the linear fittings. (e), SdH oscillations in a few-layer n-type $WSe_2$ sample. The inset is the schematic band structure of few-layer $WSe_2$. (f), Numerical derivative dR/dB plotted as a function of 1/B for data at $V_g$=70 V in (e). The inset reveals the fast Fourier transformation result of the oscillations.

|  | $\mu_{Field}$ (cm$^2$/V s) | $\mu_{Hall}$ (cm$^2$/V s) | Carrier density (cm$^{-2}$) | $\gamma_{Field}$ | $\gamma_{Hall}$ | Contact resistance (kΩ·μm) |
|---|---|---|---|---|---|---|
| MoS$_2$ | 14,000 | 9,900 | 8.8×10$^{12}$ | 2.4 | 2.3 | 0.5 |
| WSe$_2$ | 8,600 | 7,100 | 5.1×10$^{12}$ | 1.9 | 1.9 | 0.2 |
| WS$_2$ | 16,000 | 8,000 | 5.3×10$^{12}$ | 2.0 | 2.1 | 0.5 |
| MoSe$_2$ | 4,400 | 2,100 | 1.3×10$^{13}$ | 1.9 | 2.1 | 0.3 |
| MoTe$_2$ | 2,300 | 600 | 6.2×10$^{12}$ | 1.5 | 1.3 | 12 |
| P-type WSe$_2$ | 8,550 | 3,300 | 4.3×10$^{12}$ | 1.0 | 0.9 | 4.8 |

**Table 1** | Summary of the transport performance of the h-BN encapsulated few-layer TMDCs characterized at T=2 K.

# Universal low-temperature Ohmic contacts for quantum transport in transition metal dichalcogenides

Shuigang Xu[1,†], Zefei Wu[1,†], Huanhuan Lu[1,†], Yu Han[1], Gen Long[1], Xiaolong Chen[1], Tianyi Han[1], Weiguang Ye[1], Yingying Wu[1], Jiangxiazi Lin[1], Junying Shen[1], Yuan Cai[1], Yuheng He[1], Fan Zhang[2], Rolf Lortz[1], Chun Cheng[3], Ning Wang[1,*]

[1]Department of Physics and Center for 1D/2D Quantum Materials, the Hong Kong University of Science and Technology, Clear Water Bay, Hong Kong, China

[2]Departement of Physics, the University of Texas at Dallas, Richardson, Texas 75080, USA

[3]Department of Materials Science and Engineering and Shenzhen Key Laboratory of Nanoimprint Technology, South University of Science and Technology, Shenzhen 518055, China

[†]These authors contributed equally

[*]Corresponding author: phwang@ust.hk

1. **Device fabrication**

Figure S1 shows the fabrication of the h-BN/TMDC/h-BN device by the dry transferring method followed by a chemical etching processes and electron beam evaporation. A $WSe_2$ flake is used for making the device shown in Figure S1.

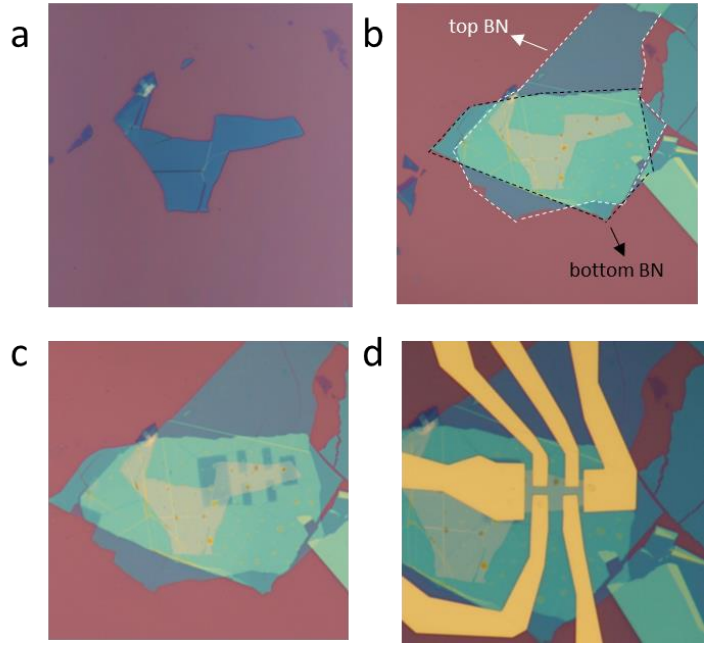

**Figure S1: Optical images of TMDC device fabrication process. a**, Few-layer $WSe_2$ flakes exfoliated on a $SiO_2$/Si substrate. **b**, A $WSe_2$ flake picked up by a h-BN sheet and transferred to another h-BN sheet. **c**, Hall bar pattern generated by standard e-beam lithography techniques and etched by the reactive ion etching process. **d**, The metal electrodes deposited on the sample via e-beam evaporation techniques.

2. **Hysteresis effects**

Figure S2 shows the positive and negative sweeps of the $G$-$V_g$ curves which are perfectly overlapped to each other, indicating that there is no hysteresis effect. The charge trapping effects have been effectively suppressed in these h-BN encapsulated structures.

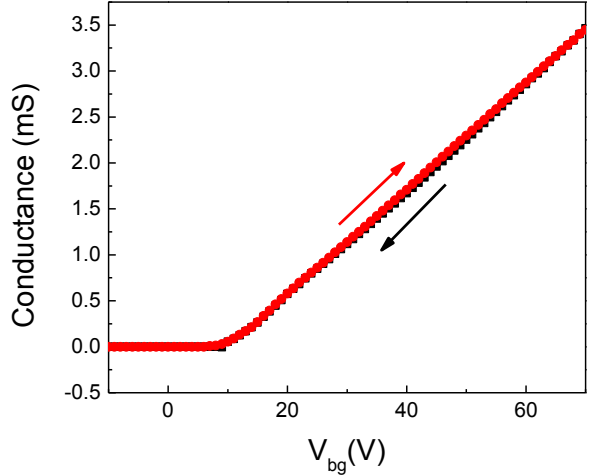

**Figure S2**: **The positive and negative sweeps of the conductance as a function of gate voltages at 2 K.**

## 3. Ohmic contacts

Ohmic contacts have been achieved in all samples in this study. Linear $I_{ds}$-$V_{ds}$ curves are observed at various temperatures as shown in Figure S3. To investigate the contact barriers in our devices, sample conductance is measured as a function of $V_g$ for different excitation voltages through a four-terminal configuration (see Figure S3c). All the curves coincide to a single curve, indicating that the injection of charge carriers is independent of the excitation voltages. Thus, the electrical properties of the channel materials are truly reflected by the transport data. Similar linear $I_{ds}$-$V_{ds}$ curves are observed in other materials as shown in Figure S4.

The barrier free contacts in our samples have been further confirmed using the Arrhenius plot[1] of $\ln(I_{ds}/T^{3/2})$ versus 1000/T. The Arrhenius plots of temperature dependent transport of n-type $WSe_2$ and $MoSe_2$ are shown in Figure S5 with $I_{ds}$=0.1 V. All the fitting curves at various gate voltages have similar positive slopes, which indicate that the contact barriers are almost independent on the back gates voltages. This kind of contacts is very different from the graphene/MoS$_2$ contact[2, 3].

We find that the h-BN encapsulated structure is very critical for the formation of reliable low-temperature Ohmic contacts. Without using the bottom or top h-BN, all TMDC devices show very poor contacts, which can be found from the $I_{ds}$-$V_{ds}$ curves obtained at room- and low-temperatures. By applying the controlled etching process (using the same contact metals (Ti/Au)), we fabricated two kinds of devices to demonstrate the functions of h-BN: Device-A without bottom h-BN and Device-B without top h-BN. Device-A displays serious non-linear behaviors at room- and low-temperatures as shown in Figure S6, indicating the impurities induced gap states at the interface between SiO$_2$ and TMDC, resulting in poor contacts. Device-B has a better performance, but asymmetrical $I_{ds}$-$V_{ds}$ curves at low-temperatures, as shown in Figure S7e, indicating that a small Schottky barrier should occur at the contacts. Therefore, we believe that the clean surface realized by h-BN encapsulation plays an important role in achieving low-temperature Ohmic contacts.

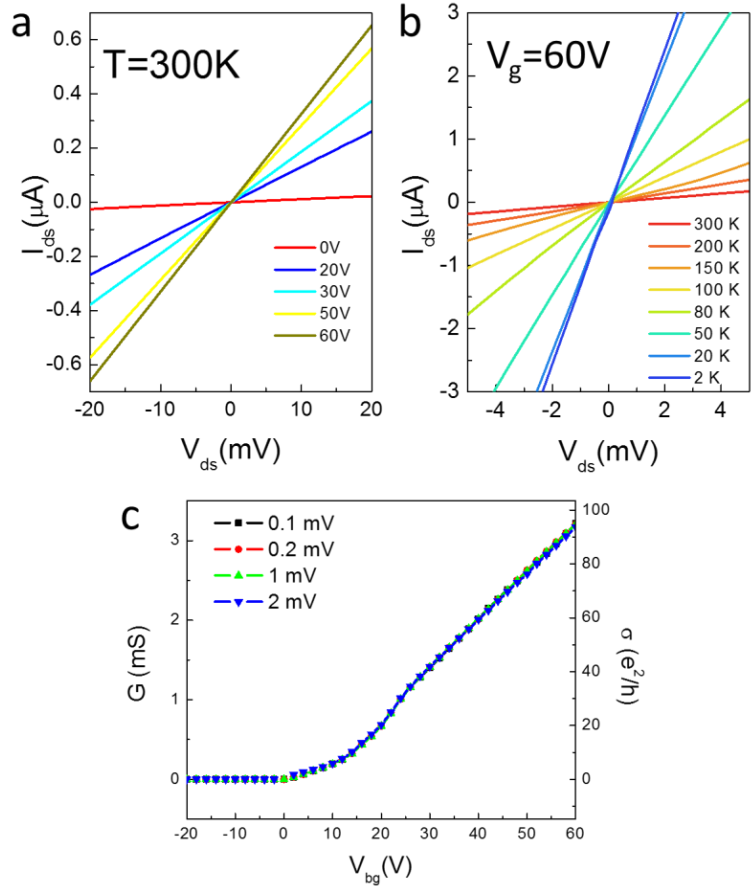

**Figure S3**: **Contact performance at different temperatures. a**, $I_{ds}$-$V_{ds}$ characteristics obtained by a two-terminal configuration at room temperature. **b**, Two-terminal output curves at various temperatures with back gate voltages kept at 60 V. Ohmic contacts are observed in a wide range of temperatures. **c**, Conductance plotted as a function of back gate voltages with different excitation voltages at 2 K.

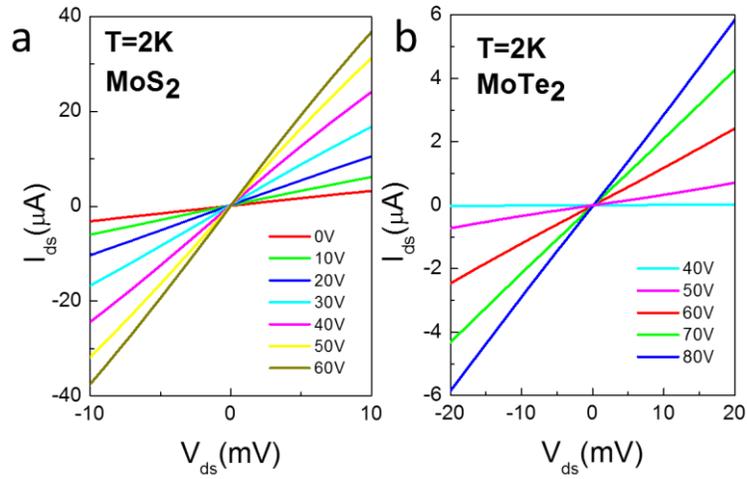

**Figure S4: Two-terminal $I_{ds}$-$V_{ds}$ characteristics at T=2 K for n-type MoS$_2$ and MoTe$_2$ devices.**

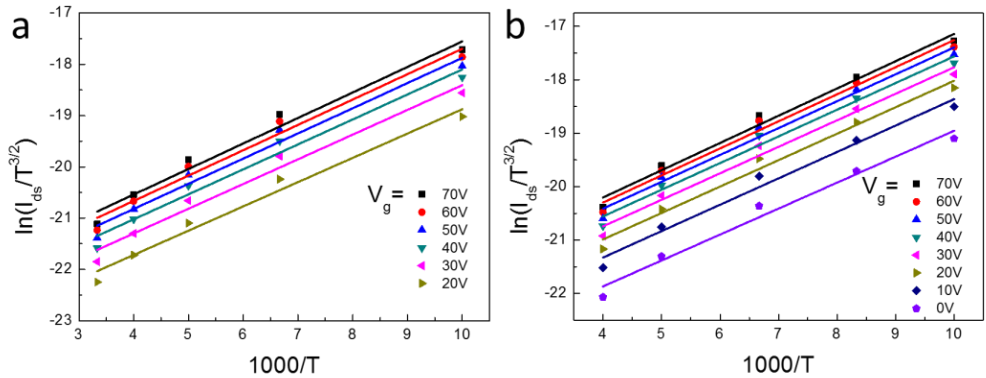

**Figure S5**: **Arrhenius plots of ln ($I_{ds}/T^{3/2}$) versus 1000/T at various back gate voltages** for WSe$_2$ (**a**) and MoSe$_2$ (**b**). The positive slope indicates zero barrier in the contacts.

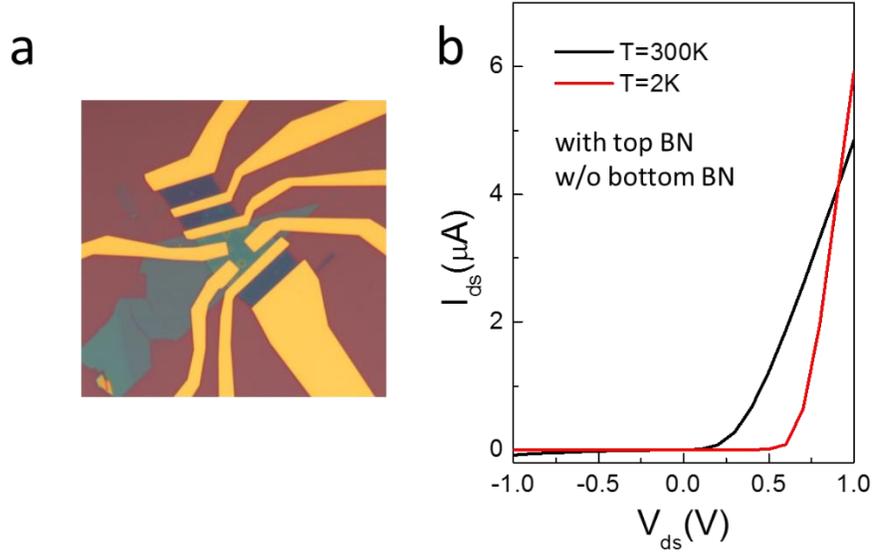

**Figure S6**: **Control experiments and the output characteristics ($I_{ds}$-$V_{ds}$ curves) of few-layer WSe$_2$ without bottom h-BN**. **a**, optical images of the device. **b**, $I_{ds}$-$V_{ds}$ curves of the electrodes fabricated on the WSe$_2$ sample protected by a top h-BN.

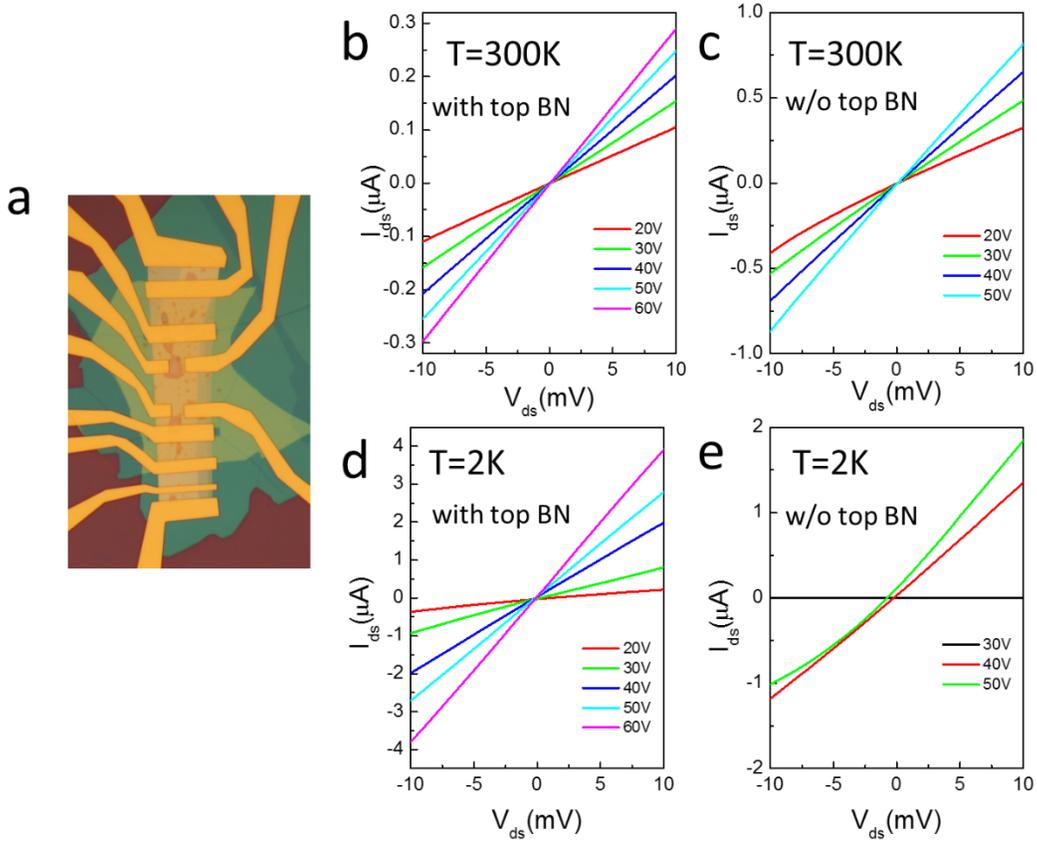

**Figure S7: Control experiments and the output characteristics ($I_{ds}$-$V_{ds}$ curves) of few-layer WSe$_2$ without a top h-BN**. **a**, an optical image of the device. **b, d**, $I_{ds}$-$V_{ds}$ curves of the electrodes fabricated on the WSe$_2$ protected by a top h-BN. **c, e**, $I_{ds}$-$V_{ds}$ curves of the electrodes fabricated on the WSe$_2$ without top h-BN protection.

## 4. Temperature dependent conductance in other TMDCs.

Figure S8 shows temperature dependent conductance at different gate voltages for various TMDC samples. With Ti as the contact metal, all of the TMDCs show n-type behaviors, except the MoTe$_2$ devices at room temperature. Because MoTe$_2$ has more narrow bandgap than other TMDCs, the Schottky barrier for the hole injection in MoTe$_2$ is smaller. Therefore, we can observe the ambipolar effect in MoTe$_2$ at room temperature with the hole conducts at negative gate voltages as shown in the inset of Figure S8c. The hole conductions dramatically decrease and are finally difficult to detect when the temperature goes down to cryogenic temperatures. This indicates at low temperature, thermal excitation energy becomes too small to overcome the Schottky barriers, which further limit the injection of holes.

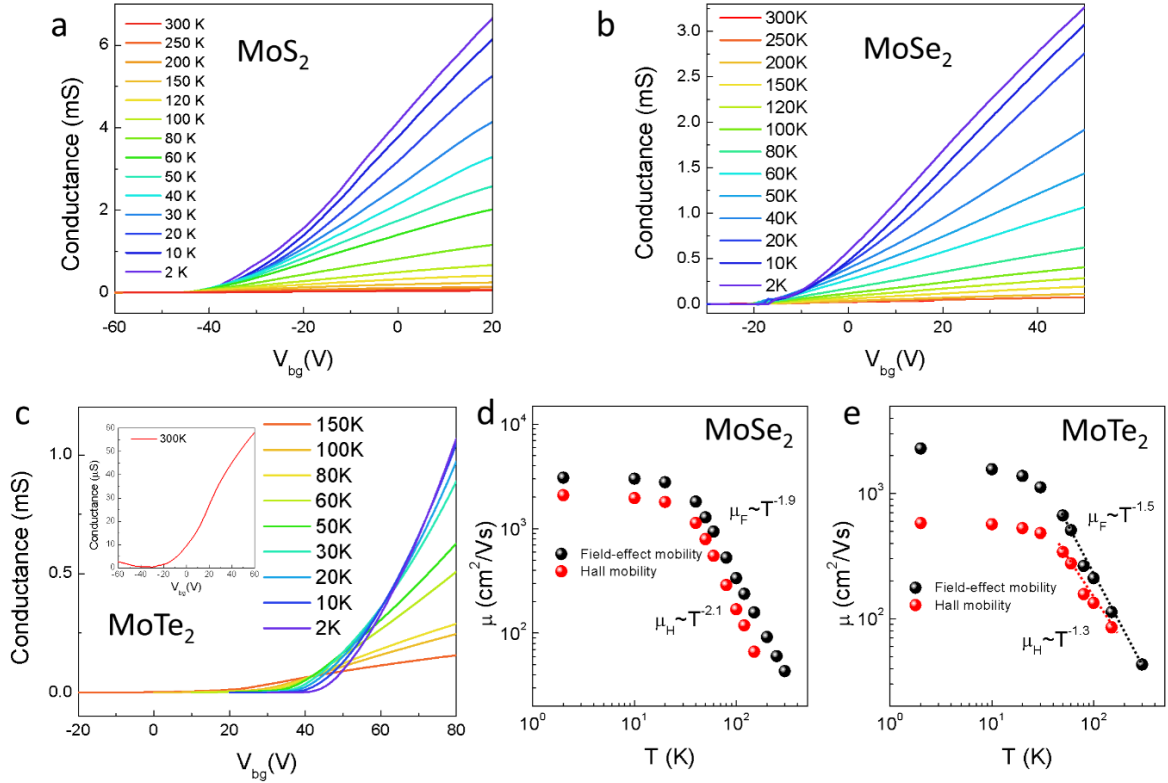

**Figure S8: The conductance changes plotted as a function of back gate voltages measured by four-terminal configurations for a, MoS₂; b, MoSe₂; c, MoTe₂.** d, e, temperature dependent field-effect and Hall mobility for (d) few-layer MoTe₂ devices and (e) few-layer MoSe₂ devices.

## 5. Low-temperature contact resistance

The contact resistance can be estimated by $R_c = R_{2p} - \frac{L_{tot}}{L_{in}} R_{4p}$, where $R_{2p}$ is two-terminal resistance, $R_{4p}$ is the four-terminal resistance, $L_{tot}$ and $L_{in}$ are the total and inner channel lengths, respectively. $R_c$ as a function of gate voltages at T=2 K are plotted in Figure S8 for different materials. The unit length contact resistance can be calculated as $R_c W/2$, considering the source and drain contributions. The results are summarized in Table 1.

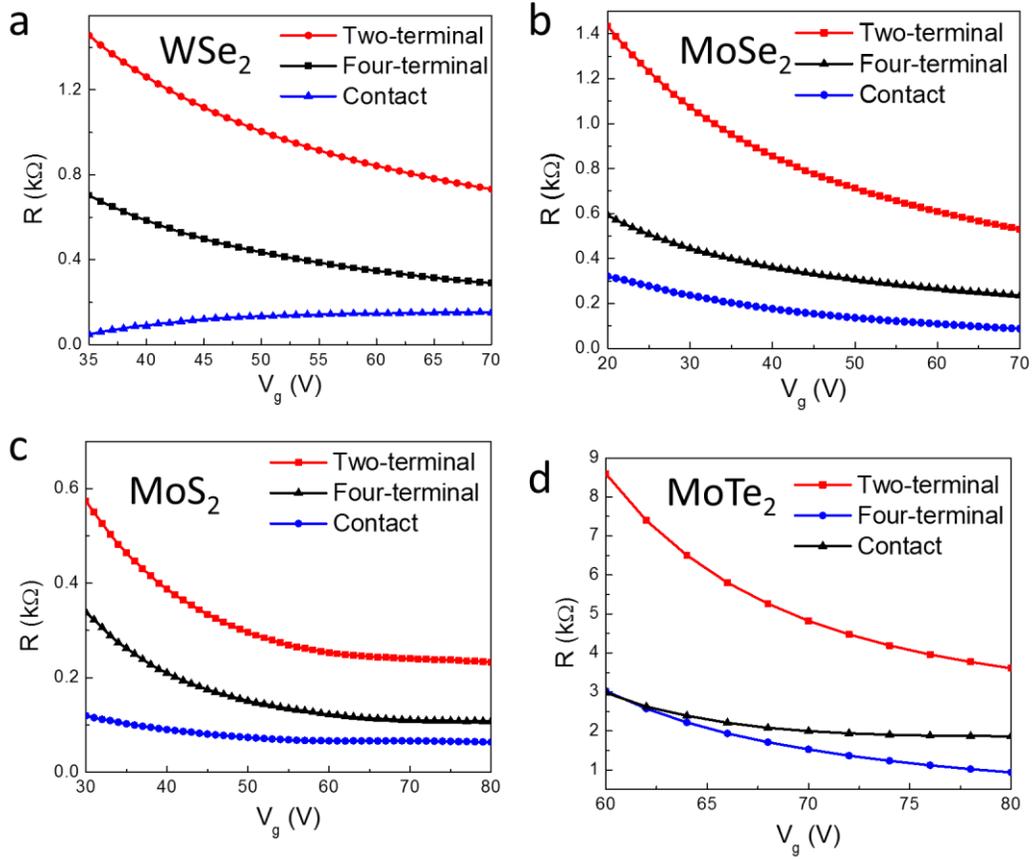

**Figure S9: Gate voltages dependent contact resistance at T=2 K for various n-type devices: a, WSe₂; b, MoSe₂; c, MoS₂; d, MoTe₂.**

6. **Carrier density and gate capacitance determined by Hall effects**

We use Hall effect data to determine the Hall mobility, density of charge carriers and gate capacitance. Figure S10a shows that the transverse Hall resistance $R_{xy}$ of the device at different $V_g$. $R_{xy}$ is linearly dependent on the magnetic field $B$. The Hall coefficient $R_H$ is extracted from the slope. Then the carrier density is calculated according to $n_{2D} = 1/R_H e$ and the Hall mobility is obtained by $\mu_H = \sigma/ne$. The variation of $n_{2D}$ as a function of $V_g$ is shown in Figure S10b. The data can be fitted by $n_{2D} = n_0 + C_g V_g/e$, whose slope yields $C_g = 1.14 \times 10^{-8}$ F/cm$^2$. This gate capacitance is used to calculate the field-effect mobility.

Figure S11 illustrates that the Hall mobility is dependent on the carrier density. The density dependence of the Hall mobility is due to the long-range Coulomb impurities in the samples. Increasing the carrier density enhances the screening of Coulomb potential, which results in an increase of the Hall mobility. The carrier density dependence of the Hall mobility leads to that the measured $\mu_H$ is smaller than $\mu_F$. Because when we substitute $\sigma = ne\mu_H$ and $n = C_g(V_g - V_{th})/e$ into the definition of field-effect mobility $\mu_F = (d\sigma/dV_g)/C_g$, we will find $\mu_F = \mu_H + nd\mu_H/dn$. Therefore, $\mu_F$ will be larger than $\mu_H$ when $d\mu_H/dn > 0$.

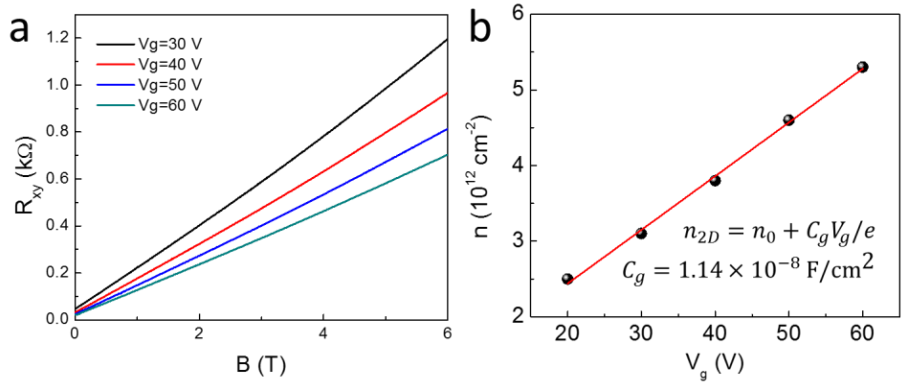

**Figure S10: Capacitance extracted from Hall effect measurements. a**, The Hall resistance plotted as a function of magnetic fields for different back gate voltages. **b**, The electron density extracted from the Hall coefficient $R_H$ at different voltages.

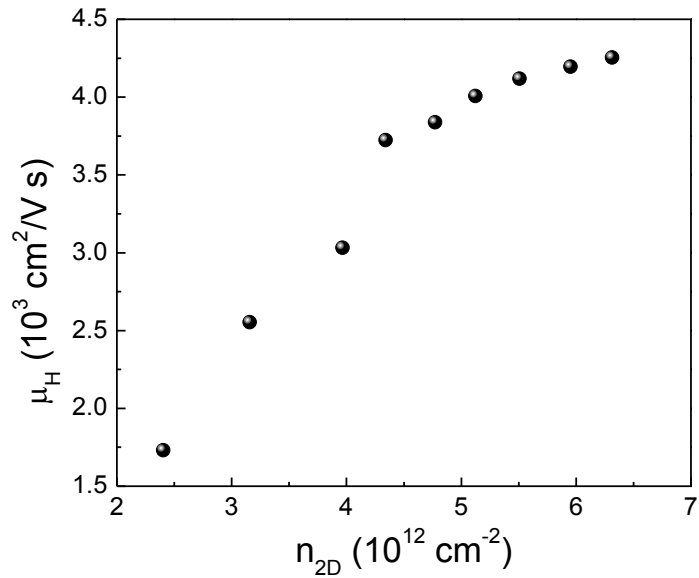

**Figure S11: The carrier density dependence of the Hall mobility in a 5L WSe$_2$ sample.**

## 7. Stability of h-BN encapsulated TMDC devices

The h-BN encapsulated devices can maintain their high performance after a long time storing in air. Figure S12 shows the mobility of a WSe$_2$ device is still over 8000 cm$^2$/V s after 4 months. The slight shift of the threshold voltage may be due to the influence of the impurity levels in SiO$_2$.

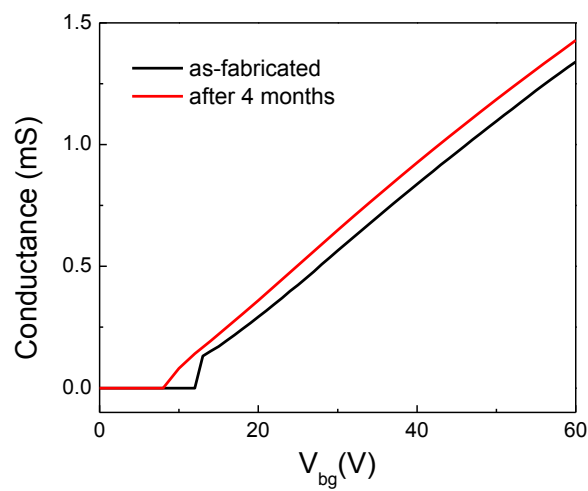

**Figure S12: The changes of the electrical transport characteristics of a WSe$_2$ sample after storing in air for 4 months.**

8. Transport behaviors of few-layer WSe$_2$ prepared on SiO$_2$/Si substrates

We have studied the transport behaviors of few-layer WSe$_2$ exfoliated on SiO$_2$/Si substrates. Two kinds of metal contacts, Ti/Au and Pd/Au have been used for comparison. As shown in Figure S13 and S14, both contacts show poor performance since large Schottky barriers formed at the interfaces between the electrodes and WSe$_2$. It is difficult to measure the conductance by four-terminal configurations or at low temperatures due to the large contact resistances. Figure S13 and S14 show the characteristics of Pd and Ti contacts (tow-terminal configuration), respectively. For Pd contacts, the measured electron mobility is only 0.2 cm$^2$/V s and hole mobility 0.04 cm$^2$/V s. For Ti contacts, the hole mobility is 8.7 cm$^2$/V s. Moreover, the devices prepared on SiO$_2$ show large hysteresis effects, indicating that there are lots of charge traps at the interface between SiO$_2$ and WSe$_2$.

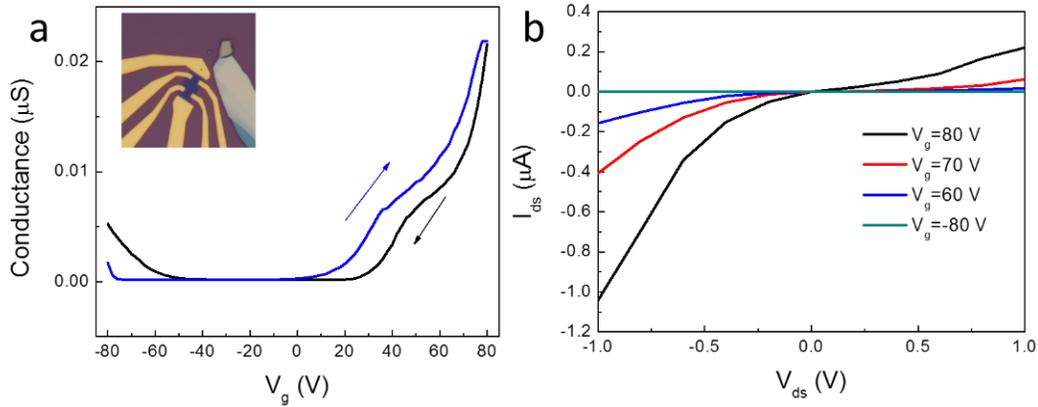

**Figure S13: Transport behaviors of a few-layer WSe$_2$ sample prepared on a SiO$_2$/Si substrate with Pd/Au electrodes**. **a**, Conductance measured as a function of gate voltages at room temperature. The inset is the optical image of the device. **b**, I$_{ds}$-V$_{ds}$ curves measured at different gate voltages.

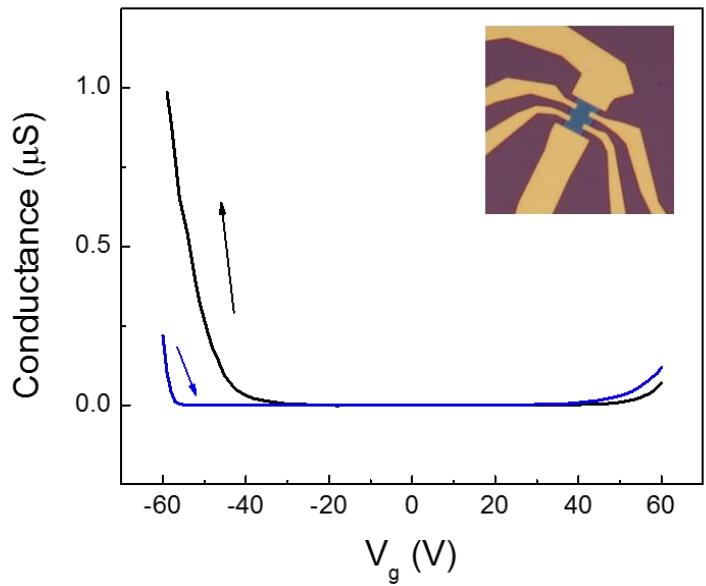

**Figure S14: Room-temperature transport behaviors of a WSe$_2$ sample prepared on a SiO$_2$/Si substrate with Ti/Au contact.**

**9. Transport behaviors of h-BN encapsulated few-layer WSe$_2$ with edge contacts.**

Although the one-dimensional edge contacts for h-BN encapsulated graphene[4] show excellent electrical performance, the same fabrication did not produce satisfactory results for h-BN encapsulated few-layer WSe$_2$. To achieve the edge contacts, 4 sccm O$_2$ and 40 sccm CHF$_3$ were used. The inset in Figure S15 shows the Hall bar pattern after the etching process. Two kinds of electrodes, Ti/Au and Pd/Au have been tested as edge contacts. Although the edge contact does work, the performance are not good enough. As shown in Figure S15, the hole mobility obtained using Pd/Au edge contacts was about 5 cm$^2$/V s. For Ti/Au edge contacts (Figure S16), we obtained the electron mobility of about 6.4 cm$^2$/V s at $V_{ds} = 0.5$ V.

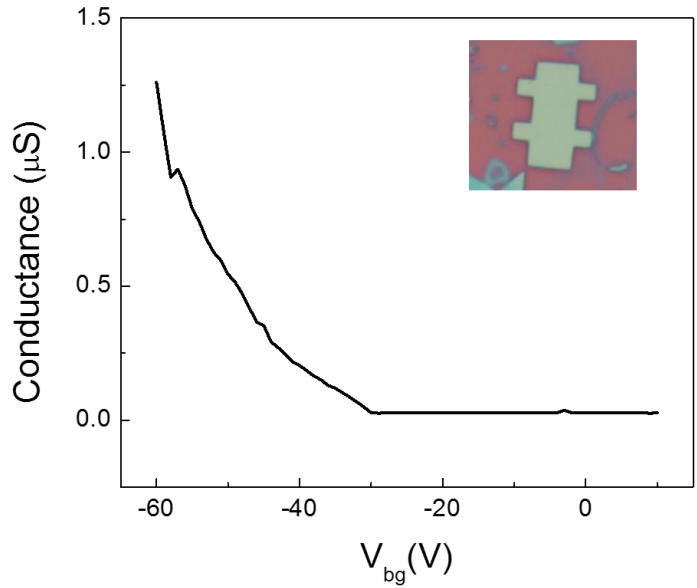

**Figure S15**: Transport behaviors of h-BN encapsulated WSe$_2$ measured with Pd/Au edge contacts at room-temperature.

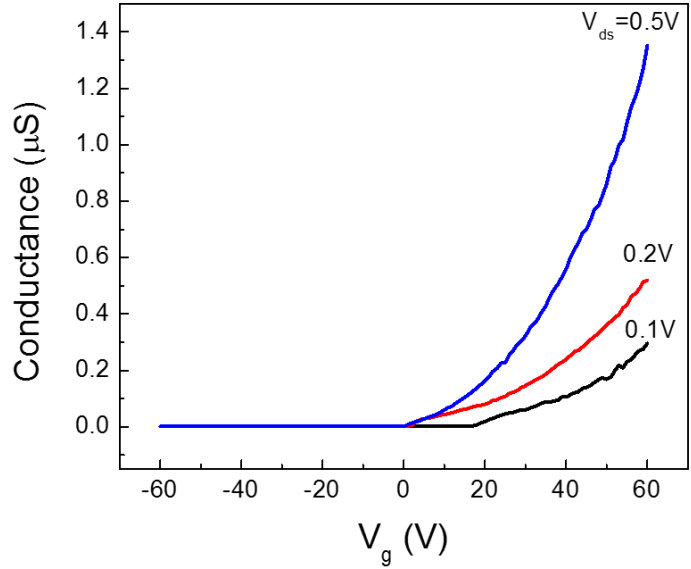

**Figure S16:** Transport behaviors of h-BN encapsulated WSe$_2$ measured with Ti/Au edge contacts at room-temperature. Excitation voltages of 0.1 V, 0.2 V and 0.5 V are used for the measurements.

## 10. Thickness determination by atomic force microscopy

We use AFM to identify the thickness of TMDC samples. The thickness of single layer WSe$_2$ is about 0.7 nm. Figure S17 shows AFM data for the 5L and 8L WSe$_2$ samples.

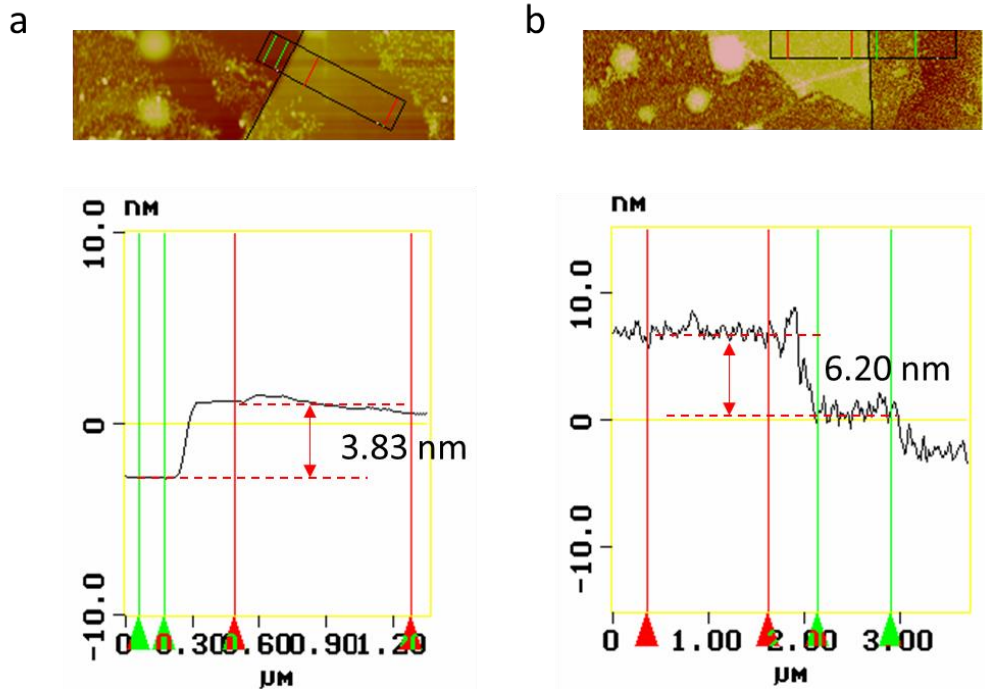

**Figure S17:** AFM results obtained from few-layer $WSe_2$ samples. **a**, 5L sample. **b**, 8L sample.

## 11. Optical characterization of few-layer $WSe_2$ and $MoSe_2$ samples

Figure S18 shows the photoluminescence (PL) and Raman scattering data taken from $WSe_2$ samples with different thicknesses. The wavelength of the excitation laser source is 514 nm. For PL, single layer $WSe_2$ exhibits a high intensity around 750 nm, indicating the direct band gap nature. For $WSe_2$ samples with thickness >2 layers, indirect gap inter-band transition appears and the total emission intensity dramatically decreases. The Raman spectra show clear band at $A_{1g}$-LA (136 cm$^{-1}$), $E^1_{2g}$ (248 cm$^{-1}$), $A_{1g}$ (258 cm$^{-1}$), in active $B^2_g$ (308 cm$^{-1}$), $2E^1_g$ (359 cm$^{-1}$), $A_{1g}$+LA (371 cm$^{-1}$), $2A_{1g}$-LA (395 cm$^{-1}$) modes, where LA stands for longitudinal acoustic single degenerated vibrations in the lattice[5, 6]. The PL and Raman spectra of $MoSe_2$ are shown in Figure S19. The direct A and B excitons of $MoSe_2$ are at 785 nm and 700 nm respectively. The peaks at 240 cm$^{-1}$ and 286 cm$^{-1}$ in the Raman spectra of MoSe$_2$ are the out-of-plane A$_{1g}$ mode and in plane E$^1_{2g}$ mode respectively.

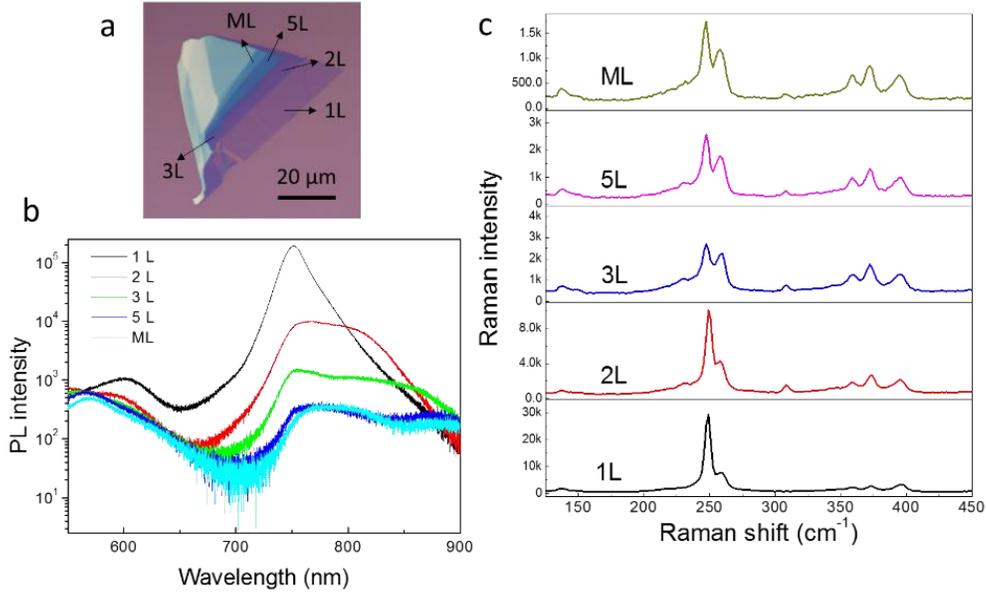

**Figure S18: Optical characterization of few-layer WSe$_2$. a**, Optical image of the WSe$_2$ flake consisted of different thickness regions. **b**, PL spectra and **c**, Raman spectra taken from different regions in the sample.

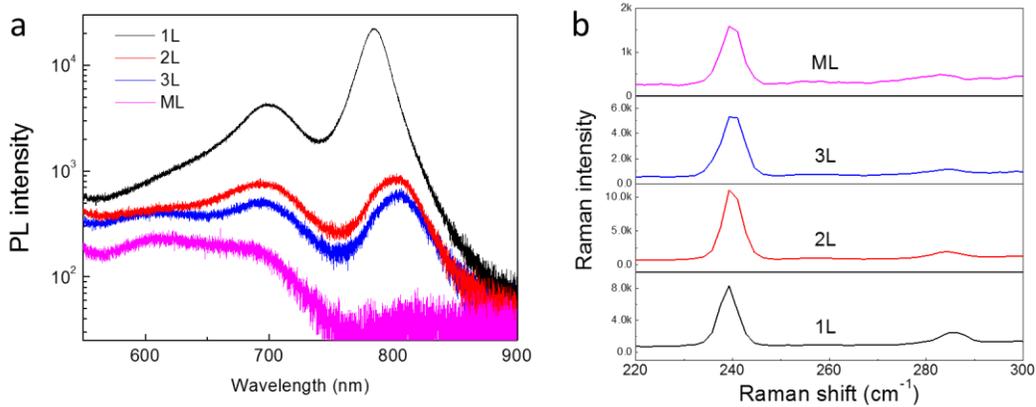

**Figure S19: Optical characterization of few-layer MoSe$_2$. a**, PL spectra and **b**, Raman spectra of different layers.

## 12. Raman characterization of controlled area etching

To identify the influence of controlled area etching on sample quality, we used Raman spectroscopy to examine the sample. The optical image of the sample is shown in Figure S20. Few-layer $WSe_2$ flake was picked up by a top h-BN and transferred to a bottom h-BN. The window of position 'P3' was opened through the controlled etching process. The Raman results show no observable change in position 'P3' compared with position 'P1' and 'P2', indicating that the sample was not significantly damaged during the controlled etching process. It's reported that the doping in TMDCs can influence the position of Raman peak[7]. We did not observe this kind of Raman shift in position 'P3', indicating that the sample was not significantly doped during the etching process.

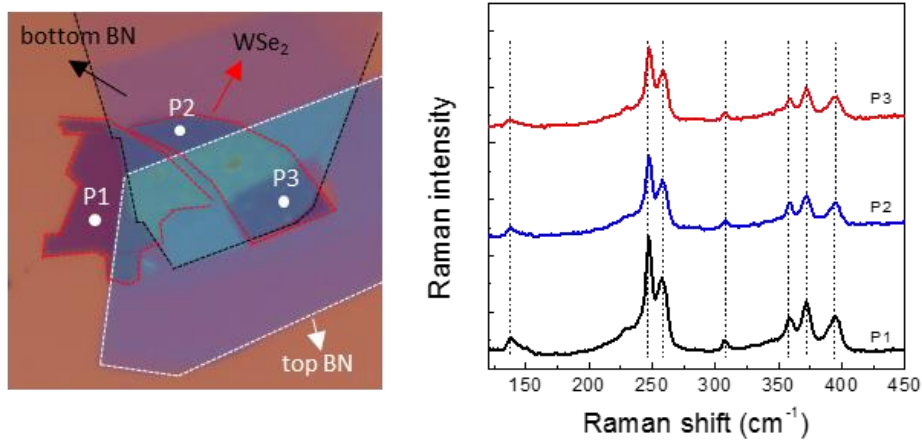

**Figure S20: Raman characterizations of few-layer $WSe_2$ after controlled etching process. a**, optical image of h-BN/$WSe_2$/h-BN sandwiched structure. Position 'P3' is opened by controlled etching process. **b**, Raman spectroscopy at different position indicated in (a).

## 13. Performance summary of h-BN encapsulated TMDC devices

| Sample | $\mu_{Field}$(T=2K) cm²/V s | $\mu_H$(T=2K) cm²/V s | $\mu_{Field}$(T=300K) cm²/V s | $\mu_H$(T=300K) cm²/V s |
|---|---|---|---|---|
| WSe₂ (8L) | 8600 | 7100 | 104 | 88 |
| WSe₂ (5L) | 6200 | 4200 | 106 | 57 |
| WSe₂ (3L) | 3900 | 1334 | 93 | 46 |
| WSe₂ (2L)* | 1200 |  | 38 |  |
| WSe₂(p-type) | 8550 | 3300 | 192 | 126 |
| MoS₂-S1 | 12700 | 7900 | 59 | 52 |
| MoS₂-S2 | 14000 | 9900 | 60 | 50 |
| MoSe₂-S1 | 4400 | 1700 | 40 | 20 |
| MoSe₂-S2 | 3100 | 2100 | 43 | 34 |
| MoTe₂-S1 | 2300 | 600 | 43 | 31 |

*\* Limited by the sample size, four-terminal device was fabricated for 2L WSe₂ sample. Therefore, the Hall mobility for this device was absent.*

**Table S1: Summary of mobility performance for different samples.** For WSe₂ samples, the layer numbers are marked. Other samples are labeled by serial numbers.

## 14. Angular dependence of the SdH oscillations

The dimensionality of the electronic states can be identified by measuring the angular dependence of the SdH oscillations. Figure S6 shows the magneto-resistance plotted as a function of magnetic fields at different tilt angles θ (between the normal axis of the substrate and B fields). By plotting the derivative dR/dB as a function of $(B\cos\theta)^{-1}$, we find the oscillation depends only on the perpendicular component of the magnetic fields, indicating the 2D nature of carriers in few-layer WSe₂.

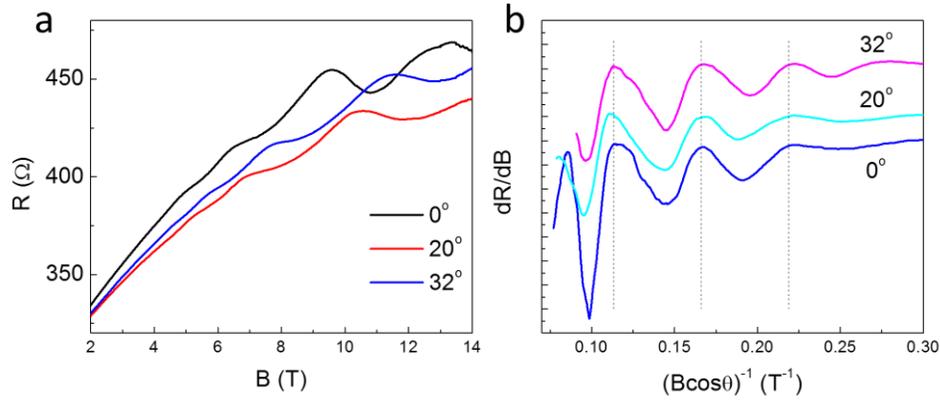

**Figure S21: Angular dependence of the quantum oscillations**. **a**, Magneto-resistance measured at different orientations with respect to the normal direction of the substrate in a n-type WSe$_2$ sample. The back gate voltage is kept at 70 V. **b**, Numerical derivative dR/dB as a function of the inverse of the component of B perpendicular to the substrate plane. An offset was added in the curves for clarity.

## 15. SdH oscillations in few-layer MoSe$_2$

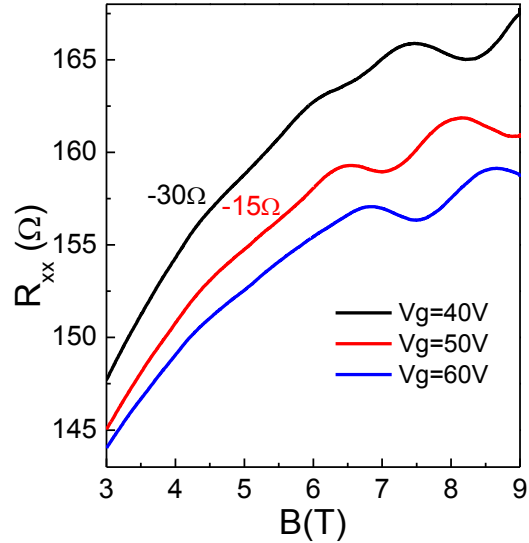

**Figure S22: SdH oscillations in a n-type few-layer MoSe$_2$ sample.** Some data were plotted by subtracting a given value for clarity.